\documentclass[11pt]{article}

\usepackage[english]{babel}

\RequirePackage{natbib}
\renewcommand\cite{\citep}	
\usepackage[a4paper,top=2cm,bottom=2cm,left=3cm,right=3cm,marginparwidth=1.75cm]{geometry}

\usepackage{amsmath}
\usepackage[multiple]{footmisc}
\usepackage{graphicx}
\usepackage{makecell}

\usepackage[dvipsnames]{xcolor}
\usepackage[colorlinks=true,allcolors=RoyalBlue]{hyperref}

\newcommand{\comment}[2]{}  
\newcommand{\adg}[1]{\comment{blue}{ADG: #1}}

\newcommand{\rasika}[1]{\comment{purple}{Rasika: #1}}

\newcommand{\jw}[1]{\comment{green}{JW: #1}}

\newcommand{\webcite}[3]{\footnote{\href{#1}{\textit{#2}}, #3.}}

\title{Co-audit: tools to help humans\\double-check AI-generated content}
\author{
Andrew D. Gordon (editor),
Carina Negreanu (editor),
José Cambronero,\\
Rasika Chakravarthy,
Ian Drosos,
Hao Fang,
Bhaskar Mitra,
Hannah Richardson,\\
Advait Sarkar,
Stephanie Simmons,
Jack Williams,
Ben Zorn\\[2ex]
Microsoft Research
}
\date{October 2023}

\begin{document}
\maketitle

\begin{abstract}
Users are increasingly being warned to check AI-generated content for correctness. Still, as LLMs (and other generative models) generate more complex output, such as summaries, tables, or code, it becomes harder for the user to audit or evaluate the output for quality or correctness. Hence, we are seeing the emergence of tool-assisted experiences to help the user double-check a piece of AI-generated content. We refer to these as co-audit tools. Co-audit tools complement prompt engineering techniques: one helps the user construct the input prompt, while the other helps them check the output response. As a specific example, this paper describes recent research on co-audit tools for spreadsheet computations powered by generative models. We explain why co-audit experiences are essential for any application of generative AI where quality is important and errors are consequential (as is common in spreadsheet computations). We propose a preliminary list of principles for co-audit, and outline research challenges.


\end{abstract}

\adg{P0=showstopper P1=nice-to-have P2=next version}

\adg{Keep in sync with \href{https://microsoft.sharepoint-df.com/:p:/t/TheGreatWFHExperiment-NFWOrganizers2/ES02V2ycUmtAntvjdZnrDtwBsyqqVpenl9dEPPm8N4DFKg?e=nlhAEP}{NFW 2023}}

\adg{Publication \href{https://www.microsoft.com/en-us/research/wp-admin/post.php?post=965220&action=edit}{page} (target of aka.ms/co-audit)}

\adg{Policy: US English spelling}
\adg{Policy: Prefer not to use "co-auditor" as it feels anthropomorphic, as discussed earlier}

\section{Introduction}
%

\subsection{Context: the rise of copilots powered by foundation models}

In this paper, we use the term \emph{copilot} for any system offering an interactive dialogue (not necessarily a textual chat) with generative AI.
%
Copilots rely on foundation models \citep{bommasani2022opportunities} to generate a response given a prompt provided explicitly or implicitly by the human user.
The response may come directly from the model, or indirectly through the use of external tools or plugins.
The prompt and response may include structured information such as lists or tables, code in a programming language, or actions to be executed.
In the simplest case of a large language model (LLM), the prompt and response are textual, but in general foundation models may process other modalities such as audio, images, or video.

In the past two years, many copilots have become available as commercial tools.
%
%
In mid 2021, GitHub Copilot was released, aimed at software developers.
In late 2022, OpenAI launched ChatGPT, a general-purpose tool for the general public.
In early 2023, to list just a few prominent examples, Microsoft released Bing Chat\webcite{https://www.nytimes.com/2023/02/07/technology/microsoft-ai-chatgpt-bing.html}{A Tech Race Begins as Microsoft Adds A.I. to Its Search Engine}{7 February 2023}
and announced Microsoft 365 Copilot,\webcite{https://blogs.microsoft.com/blog/2023/03/16/introducing-microsoft-365-copilot-your-copilot-for-work/}{Introducing Microsoft 365 Copilot — your copilot for work}{16 March 2023} Google released its Bard chatbot,\webcite{https://www.cnbc.com/2023/03/21/google-ceo-pichai-memo-to-employees-on-bard-ai-things-will-go-wrong.html}{Google CEO tells employees that 80,000 of them helped test Bard A.I.}{21 March 2023} and OpenAI released its Code Interpreter,\webcite{https://openai.com/blog/chatgpt-plugins}{ChatGPT plugins}{23 March 2023} one of several plugins for ChatGPT.
%
Meanwhile, researchers have explored copilots, such as Sensecape \cite{DBLP:journals/corr/abs-2305-11483}, an interface that provides multilevel abstraction and sensemaking, that illustrate how dialogue with a copilot can go beyond textual chat.
Copilots promise to transform the practice of knowledge work, while triggering public debate about its future.

\subsection{The prompt-response-audit cycle}\label{sec:cycle}

\begin{figure}
\begin{center}
\includegraphics[width=\textwidth]{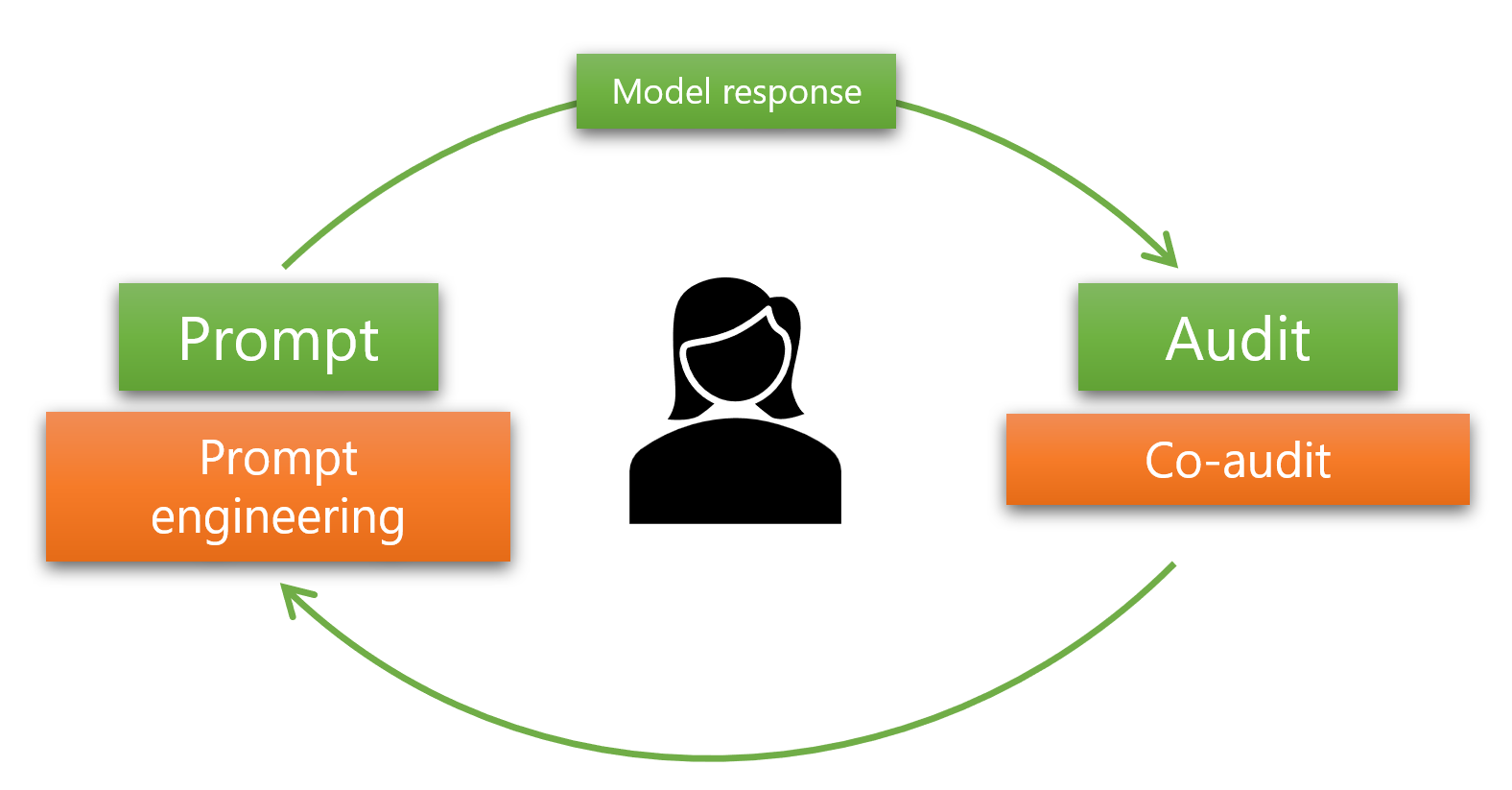} 
\end{center}
\caption{The \emph{prompt-response-audit} cycle.}\label{fig:cycle}
\end{figure}

From the perspective of the human user, the interactive dialogue with the copilot consists of a \emph{prompt-response-audit} cycle, as shown in Figure~\ref{fig:cycle}.
\begin{enumerate}
\item Prompt.
Given an \emph{intent} or purpose, the user prepares a \emph{prompt} to pass to the copilot.
The user may rely on prompt engineering skills they have learnt, or on prompt engineering tools that help to build the prompt \cite{liu2021pretrain,ZamfirescuPereira2023WhyJC,nam2023inide}.
\item Response.
The user waits while, given the prompt, the copilot generates a \emph{model response}.
The response may be the direct output from a generative model, or it may be the result of an interaction between a generative model and a set of external tools or plugins.
Often, the copilot warns the user that the response may contain errors and should be checked.
\item Audit.
Given their intent, the user evaluates the response, and decides to accept it, or to iterate further in the cycle.
We call this human evaluation of the response an \emph{audit}, in the sense of a systematic review.
Depending on the intent, audit may be to an objective standard, or subjective to the user.
During the audit, the user may decide one or more of the following.
\begin{enumerate}
\item[(A)] They are satisfied with the response, given the intent.
\item[(B)] There is a mistake in the response, given the intent.
\item[(C)] The intent itself was mistaken, and needs to be refined.
\end{enumerate}
In case (A), the cycle for this intent ends.
In case (B), the user may directly repair the mistake, or do so by continuing with an updated prompt.
In case (C), the user continues with an updated intent and prompt, or may directly repair the mistake.
\end{enumerate}


In this context, audit is the systematic review of a piece of AI-generated content.

We introduce the term \emph{co-audit} for any tool-assisted human experience for audit.

The motivation for co-audit is that the human task of auditing is hard, especially if done well, and getting harder as responses become more complex.
Co-audit encourages and assists the user to heed the warnings, and do a better job of auditing the response.
The ``co'' in co-audit is to emphasise that the human works with some tool support to do the audit.
The tool support may or may not itself use AI technology.
Co-audit helps the user double-check the response.
Co-audit need not have the rigor of formal verification tools.
Co-audit is about understanding the response, but also about the process of clarifying the intent, and effectively communicating that updated intent to the model.

Figure~\ref{fig:cycle} depicts prompt engineering and co-audit as complementary aspects of human-AI dialogue.
The one helps construct the input prompt, while the other helps double-check the output response.
We intend the cycle to apply broadly to many different situations, including textual chat between a human and an AI tool, but also mouse-based interactions in graphical interfaces where the prompt is constructed by the system from a mouse click (a form of prompt engineering).
Dually, the experience abstracted by the co-audit box in Figure~\ref{fig:cycle} may involve various sorts of interaction with the system, text-based or mouse-based, for example.

\adg{P2 restore the autosuggest example, if we consider co-audit for AI systems in general}

To give a concrete example of co-audit from the research literature, consider
ChatProtect\webcite{https://chatprotect.ai/}{ChatProtect detects and removes hallucinations of LLMs when you chat with them}{9 September 2023} \cite{mündler2023selfcontradictory}, a chat experience with a language model, enhanced with features to detect and remove hallucinations. ChatProtect detects hallucination by sampling multiple completions from the model, and testing when they disagree. The co-audit experience lets the user inspect different sentences in the response to see the effect of ChatProtect.

\subsection{The purpose of co-audit is to identify mistakes and repair them}
When auditing, the user is looking for any \emph{mistake}: any part of the response that doesn't match their intent.

Mistakes include textual hallucinations, defined by \citet{DBLP:journals/csur/JiLFYSXIBMF23} as ``generated content that is nonsensical or unfaithful to the provided source content''.
Other examples (which may or may not count as hallucinations) include a summary that omits important information, faulty claims about the world, a faulty fictional narrative that mixes up characters or timelines, faulty mathematical reasoning (including faulty arithmetic or algebra), faulty code, or a faulty image of a human hand with six fingers.

There are several concrete examples where copilots make mistakes of this sort.
For instance, \citet{bubeck2023sparks} give examples of faulty logical and arithmetic reasoning by GPT-4 and other language models.
OpenAI's Code Interpreter is effective at automating data science tasks \citep{cheng2023gpt4}.
\citet{tu2023data} report that given a data science exam, it performed well, but not perfectly, scoring 104 marks out of a possible 116.
Still, users need to be careful: a commentator noted that ``Code Interpreter made several logical mistakes that only an expert could have caught.''\webcite{https://twitter.com/svpino/status/1679135894912024576}{I rewrote around 1,000+ using OpenAI's Code Interpreter and am impressed!}{12 July 2023}
In the domain of mathematics, \citet{collins2023evaluating} develop a framework to evaluate how effectively humans can develop proofs despite the faulty algebraic reasoning made by language models.
One of their conclusions for any mathematician using an LLM is to ``pay attention'': LLMs can generate extremely compelling mathematical language and still be wrong.



The user faces three challenges when auditing the response.
These are intensifying as more knowledge workflows increasingly incorporate AI-generated content.


\begin{enumerate}

\item
It is hard to entirely eliminate mistakes at the model response step.

Despite much research \cite{lee-etal-2022-factuality,li2023inferencetime,chuang2023dola} and benchmarks \citep{gao2023enabling} hallucinations still occur. Copilots built on natural language generation (and indeed other modalities such as images) therefore inherit the possibility of hallucinations and other kinds of error.

\item
The patterns of mistakes from humans and AI are different.

The shift from human-generated to AI-generated content creates a qualitative shift in the user experience of audit. For example, checking a human-authored document summary for correctness is done with a model of the author’s intent and kinds of quality issues that are known to arise when people write text. Is the author well-meaning or adversarial? Are we looking for unsupported statements, irrelevant statements, omissions, or overemphasis? In contrast, AI-generated content has errors for which it is impossible to ascribe a “motive”. Due to hallucinations, there are broad issues of factual correctness and grounding in the source data that are more common and important for AI-generated content than for human-generated content.





\item
The user needs decision support after finding a mistake in the response. 



A mistaken response might be viewed qualitatively as an error in the system's interpretation or understanding of the intent, or as an error in executing the interpretation. Depending on how the mistake is interpreted, the user may need to update their prompt, attempt to work with and refine the mistaken output, or abandon the generative tool altogether for a particular task.

Intent matching is the user challenge of expressing their intent in a way that the model would ``understand'' correctly, and then checking whether it has done so (generalising the notion of ``abstraction matching'' from \citet{DBLP:conf/chi/LiuSNZ0T023}). When knowledge work is dominated by primarily human-generated content, intent matching is externalised to the relationship between individuals; if I ask you to do a task, and it appears that you had misunderstood me, I will take it up with you, not the software. When an individual works alone, the issue of intent matching is externalised to the user’s own conception of their skills (“am I applying this feature correctly to achieve what I want?”). But generative AI tools create a new expectation, that the user can express their intent in the way they might express it to a colleague. This creates a new type of relationship between user and system in which the responsibility for intent matching cannot fall either entirely on the user, or be externalised to the relationship with a human colleague. 


The research field of interactive machine learning commonly frames the role of the user interface (and AI ``explanations'' more specifically) as a decision support mechanism; i.e., giving the user the information they need to take the next step in refining and improving their model \cite{nunes2017systematic,sarkar2022explainable}. Co-audit systems need to incorporate decision support specifically geared towards repairing issues with intent matching and correctness, a type of support that is only necessary in a knowledge workflow dominated by AI-generated content.
\end{enumerate}

\subsection{Co-audit matters in some application areas more than others}

Co-audit experiences matter most in application areas where correctness is important and factual or reasoning errors are consequential.

\adg{P2 find examples of tools to help co-audit output from GitHub Copilot - eg at CHI?}

Three such example areas are technology, healthcare, and finance.
Table \ref{tab:areas} shows examples of tasks within each that could benefit from co-audit experiences.
The table mentions three examples of AI systems that could potentially benefit from co-audit: GitHub Copilot, Nuance\webcite{https://whatsnext.nuance.com/healthcare-ai/how-ai-automation-adds-efficiency-to-radiology-workflow/}{How can AI unlock next-generation radiology reporting?}{15 August 2022}\:\webcite{https://news.nuance.com/2023-03-20-Nuance-and-Microsoft-Announce-the-First-Fully-AI-Automated-Clinical-Documentation-Application-for-Healthcare}{Nuance and Microsoft Announce the First Fully AI-Automated Clinical Documentation Application for Healthcare}{20 March 2023}, and systems from Bloomberg. The tasks are not necessarily applicable to all the systems mentioned.

\begin{table}[h!]
  \begin{center}
    \begin{tabular}{c|c|l} 
      \textbf{Area} & \textbf{Example AI Systems} & \textbf{Example Tasks}\\
            \hline

      Technology & GitHub Copilot & project planning \cite{AutoScrum} \\
       & & code authoring \cite{CodeLLama}\\
       & & documentation \cite{code_doc}\\
       & & code reviews \cite{ACR}\\
       & & analysis \cite{InsightPilot}\\
       \hline
      Healthcare & Nuance & diagnosis \cite{Xv2023CanCA}\\
       & & prescription calculations \\
       & & authoring clinical notes \\
       & & \quad \cite{Rheumatology} \\
       & & nutrition recommendations \cite{nutrition}\\
       & & mental health \cite{mental_health}\\
       & & on the go recommendations \\
       \hline 
       Finance & Bloomberg GPT  & financial research \cite{finresearch} \\
       & \quad \cite{bloombergGPT} & trend analysis \cite{FinVis} \\
       & FinGPT & insurance policy recommendations \\
       & \quad \cite{FinGPT} & financial investments \cite{fininvest} \\
       & & business plans \cite{ProcessGPT}
    \end{tabular}
  \end{center}
      \caption{Important areas for co-audit. Specialised models and copilots have started to emerge for technology, healthcare and financial applications. There is a need and opportunity for co-audit tools in key tasks, such as the ones presented in this table.}
    \label{tab:areas}
\end{table}

The need for co-auditing for exploratory or creative areas, such as creative writing or music recommendations, is less clear as there is no definite right answer, and it is subjective whether part of the response is a mistake. Still, a critique that identifies shortcomings is valuable as a co-audit experience. If an LLM helps to write a novel, a co-audit experience could help identify shortcomings in the story or writing, and to identify potential repairs.

\adg{P2 Make sense of this: The role of tool support in such a scenario, rather than co-audit, could be to provide the user an estimate of the system's diversity and encourage the user to reassess their asks or priorities.}

Hence, the need for co-audit varies depending on properties of the users, the domain, and the task, making co-auditing a challenging and highly custom research space. There is not likely to be a single co-audit experience that dominates all application areas. Yet it seems plausible that certain principles of design for co-audit experiences may be broadly applicable across different domains.









\subsection{Co-audit matters to some people more than others}

As we design and critique co-audit mechanisms, we should bear in mind that using generative AI demands deep cognitive effort.
\citet{KT23} consider thinking, evaluating, and adapting as the crucial aspects to the user's cognitive effort as they work with generative AI.
Their work is informed by the psychological concept of metacognition (thinking about thinking) \citep{metacognition}.
\begin{enumerate}
\item Thinking: The user needs self-awareness of their concrete goals and intentions, and the ability to express their goals as effective prompts or commands.

\item Evaluating: The user needs the ability to evaluate the output, and the confidence to challenge it when it appears wrong. 

\item Adapting: The user needs flexibility either to repeat and to adapt prompts, or to correct the output.
\end{enumerate}
Thinking is the effort during the prompt part of the prompt-response-cycle in Figure~\ref{fig:cycle}, whereas evaluating and adapting are the effort when using a co-audit tool.
Evaluating outputs from generative AI is far more important and effortful compared to, for example, mere word or phrase suggestions from auto-complete powered by previous generations of machine learning technology.
The user's confidence in themselves is also important: studies of decision-making with AI (though not with generative AI) find that users' confidence in themselves may be more important than their confidence in the AI itself \citep{human-confidence-in-ai-2022,SK23,ai-assisted-decision-making}.


Moreover, anecdotal evidence suggests that users with low confidence may perform less with generative AI and even be intimidated by it.
Michael Hilton, reporting his experience introducing a code-generating language model to his software engineering class, said: “I'm very concerned that the use of these tools will further exacerbate the disparity between high and low self-efficacy students.”\webcite{https://twitter.com/michaelhilton/status/1646315922947284997}{We used a class in Foundations of Software Engineering to collaboratively learn about ChatGPT}{13 April 2023} He observed that self-confident students would decisively accept or reject AI suggestions, while less confident students would be intimidated and get stuck trying to understand bad code suggestions. The metacognitive challenges of identifying when to apply generative AI and how to avoid the trap of constantly evaluating suggestions have also been observed in professional software engineers \cite{sarkar2022programmingai}.

Hence, a challenge and an opportunity for co-audit mechanisms is to equip less intrinsically confident users to effectively and efficiently assess and incorporate suggestions from generative AI.


\subsection{This paper and its context}

The contributions of this paper are to identify co-audit experiences as an important aspect of interactions between humans and copilots, and to begin a systematic study of co-audit, including needs, case studies, some guiding principles, risks, and research questions.

This paper came about as a result of a one-day internal workshop at Microsoft in May 2023, in which all the authors participated.


We organise the rest of the paper as follows.
Section~\ref{sec:taxonomy} presents a taxonomy of the needs for co-audit.
Section~\ref{sec:spreadsheets} describes work on co-audit mechanisms for computations in spreadsheets, as a case study.
Section~\ref{sec:principles} lists general principles for co-audit designs.
Section~\ref{sec:pitfalls} describes some potential risks from co-audit mechanisms themselves, and discusses potential research on mitigations.
Section~\ref{sec:related} describes related work, and Section~\ref{sec:conc} concludes.

\section{Taxonomy of co-audit needs}\label{sec:taxonomy}

\rasika{RESOLVED. I understand this is not meant to be exhaustive but the taxonomy seems to address only the "Correctness" challenge. Should there be say a section like "Confirmation" or "Prevention" where we have ability to examine how easy it is to confirm understanding of intent and allow modification of the current understanding. Andy: re understanding of intent, we will have a guideline on that.}

The design requirements for co-audit systems likely vary substantially across different task contexts. To make progress in accounting for the diversity of human-AI collaboration scenarios, we propose a preliminary taxonomy that characterizes key dimensions along which co-audit needs differ. This taxonomy aims to delineate situational factors that drive differences in how co-audit should be conducted, such as the frequency and costs of errors, constraints on acceptable solutions, end-user expertise, and the positioning of the AI. By delineating a space of co-audit situations, this taxonomy begins to outline a framework for reasoning about the design of human-AI oversight mechanisms that are well-matched to specific work contexts. While preliminary, this taxonomy aims to illustrate the kind of conceptual grounding needed for the development of co-audit systems.

\begin{description}
\item[Ease of detection:] how easy is it for the user to detect an error in the AI output? This can range from very easy (obvious at a glance) to hard (e.g., needs careful inspection of the output, cross-referencing with other sources, etc.). For instance, correctness of an AI-generated image usually can be told at a glance. On the other hand, correctness of several hundred lines of Python performing data analysis may be far from obvious.

\adg{P2 overlap with subsection on important areas for co-audit in the intro}
\item[Cost of error:] how costly is an error in this situation? The cost can range from low (the error is inconsequential) to high (catastrophic). Automatically assessing the cost is a difficult task for tools as it depends on rich context. For example, AI introducing a formula error can be relatively inconsequential if the formula is being used as part of an exploratory analysis with no downstream uses, but the same kind of error can be catastrophic if its result is used to make a critical business decision. 

\item[Frequency of audit:] how often is co-audit required or needed in the workflow? If the aim is to provide spot checking of every transient AI-generated output, the design of co-audit tools will need to be lightweight and impose low cognitive and information processing burdens, focusing only on key aspects of intelligibility (akin to spelling and grammar checking). If errors are more infrequent, or if co-audit only occurs at significant milestones (e.g., at the end of the preparation of a document, or for formal auditing procedures) then co-audit tools can be more comprehensive and information dense. 

\item[Solution definition and constraints:] how large and well-defined is the solution space? A solution space may be very large, yet still well-defined: for example, if a LLM is asked to generate a sorting algorithm, there are a potentially infinite number of correct solutions, but it can be established definitively whether a given solution is correct. In some cases the solution may take a very wide range of acceptable forms, but which are judged subjectively (e.g., creating a nice looking design for a presentation, or writing an essay). On the other extreme, solutions may have very narrow or even singular objectively correct form (e.g., the items in a budget are accounted for correctly in a grand total).

\item[End-user expertise:] what kinds of skills and expertise does the intended end-user have? For instance, co-audit for a code-generating copilot would vary depending on whether the end-user is a professional programmer, versus a non-expert end user. Analogous situations arise in other application domains: in healthcare, for instance, a system for supporting treatment and diagnosis would present different information (and different audit challenges) for doctors seeking to make a decision grounded in clinical practice, versus patients with limited medical expertise. 

\item[Positioning:] What is the implied metaphorical relationship between the AI copilot and the human user? For example, the pair programming framing in GitHub Copilot implies co-working towards a shared goal, with tasks delegated from the user to Copilot. In contrast, meeting coach in Microsoft Teams or presenter coach in Microsoft PowerPoint position themselves as reflective critics of the users’ behavior. Different AI “personas” require different kinds of audit support. For a co-working AI like GitHub Copilot, the audit process would likely need to be focused on ensuring that the AI is enhancing the performance and skillset of the user---providing efficient, correct and relevant suggestions. For an AI with a critical role such as a coach in Teams or PowerPoint, the audit mechanism might focus more on evaluating its ability to provide constructive and accurate feedback. In any case, co-audit experiences should help assess how well the AI is fulfilling its implied metaphorical role---be it as a helper, an advisor, or a collaborator---and whether it is contributing positively to the user's overall goal.
\item[Scale:] how can we design co-audit to scale to large amounts of AI-generated content? Leveraging AI to generate content empowers individuals to be more productive.  As a result, AI will likely allow more content to be generated more quickly, resulting in the challenge of having individuals co-audit all that content.  Co-audit tools must be designed to scale in the amount of human effort as the amount of generated content increases.  Co-audit tools can benefit from existing experiences based on tools used in software development. For example, refactoring tools help developers make repeated changes throughout a code base with less effort.

\item[Collaboration:] how do we design co-audit to support collaboration?
Section~\ref{sec:cycle} frames audit as an activity by a single human using an AI tool, and co-audit as tool assistance for audit.
More generally, two or more humans could collaborate during the audit step.
A co-audit tool could enlist the help of a co-worker to help inspect or repair the response from the AI.
The co-worker could be a local colleague or a remote gig worker.
For instance, if applied to consider AI-generated content, the Find-Fix-Verify crowd programming pattern \cite{soylent2015}, which enlists gig workers to help within a productivity tool, would be an example of co-audit.
\end{description}

\section{Case study: co-audit tools for spreadsheet computations}\label{sec:spreadsheets}
We describe two related tools built at Microsoft Research as part of an internal project to synthesise spreadsheet computations from natural language. The tools shared infrastructure to generate Python using the OpenAI Codex model and to execute it using an addin for Microsoft Excel. Both these tools have co-audit experiences intended to help the end user double-check the generated code.

\subsection{Grounded abstraction matching}

Grounded abstraction matching is a technique to help users learn how to effectively communicate their intent to AI systems like code-generating large language models \cite{DBLP:conf/chi/LiuSNZ0T023}. It addresses the abstraction matching problem, where users struggle to find the right level of abstraction when expressing their goals in natural language that the LLM can reliably interpret. 

Grounded abstraction matching works by taking the user's initial natural language query, generating code from it via the LLM, and then mapping that code back into a natural language utterance that serves as an editable example of how to invoke the same behavior from the system. Concretely, the user enters a query like ``calculate average monthly sales.'' The system generates Python code from this query, e.g., \texttt{df[`monthly\_sales\_avg'] = df[`sales'] / 12}. Then it translates this code back into a grounded utterance:

\begin{enumerate}
    \item Create column \texttt{monthly\_sales\_avg}
    \item Column \texttt{sales} divided by 12
\end{enumerate}

This grounded utterance represents the level of abstraction---vocabulary, structure, specificity---needed for the system to reliably reproduce the same output. The user can edit the steps and observe the effects on the code produced, developing a mental model of how to effectively prompt the system. 


Grounded abstraction matching exemplifies co-audit by assisting users in evaluating AI-generated content in multiple ways. For example, the grounded utterance reveals how the system interpreted the original query, making mismatches in intent clear. Users can update the utterance to align intent. The step-by-step breakdown of the Python code into naturalistic utterances exposes the reasoning chain, helping non-expert users spot faulty logic. Finally, grounded utterances provide an editable target to iteratively refine queries, which gives users the decision support then need to proceed after they have received the model output.




In summary, grounded abstraction matching helps users learn how to effectively communicate with LLMs for code generation. By grounding output code back into editable, step-by-step natural language utterances, it provides a mechanism for users to co-audit the system's interpretation of their intent, spot errors, and iteratively repair issues, while exposing them to reliable patterns of interaction. 


\subsection{An end-user spreadsheet inspection tool for AI-generated code}
\begin{figure}
    \centering
    \includegraphics[width=\textwidth]{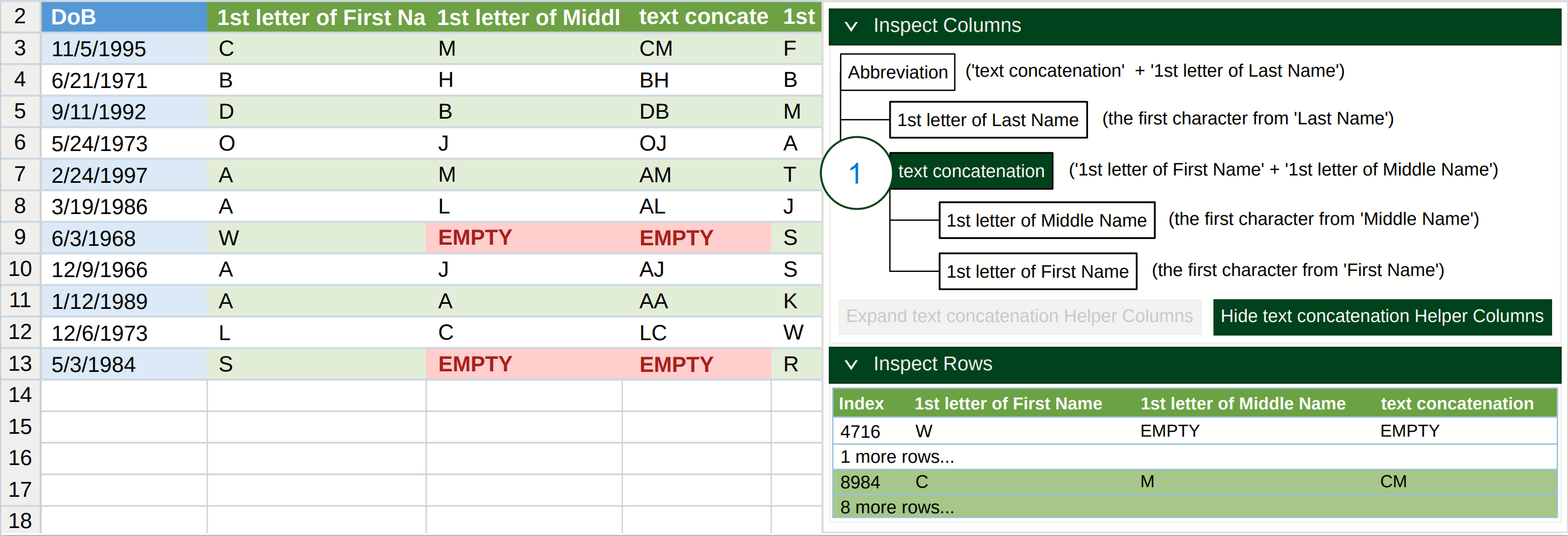}
    \caption{ColDeco}
    \label{fig:coldeco}
\end{figure}

ColDeco~\cite{ColDeco} is a spreadsheet tool for inspecting and verifying calculated calculated columns without requiring the user to view the underlying code. With generative AI, natural language is now a programming language, meaning users without coding experience can generate complex logic. ColDeco uses \emph{helper columns} and \emph{row summaries} to assist users in verifying that the program behaves correctly, or to localize the fault if it does not.

Figure~\ref{fig:coldeco} presents ColDeco as a user debugs a generated program intended to calculate an abbreviation composed of the initial letter from the  first, middle, and last names. 

The \emph{inspect columns} view displays the generated solution which has been decomposed into helper columns by the user; each generated column has a natural language description of the code associated with the column. The columns are arranged in a tree view representing the underlying program structure, with Label~1 annotating the intermediate column \emph{text concatenation}. The helper column combines the initial letters from the first and middle names, and by expanding the column in ColDeco, the new columns are added to the table.

The \emph{inspect rows} view clusters rows in the table according to evaluation behaviours that are determined through dataflow analysis. There are two clusters for this table: the rows that return a string for the abbreviation, and the rows that erroneously calculate a missing value.

Like grounded abstraction matching, ColDeco satisfies co-audit needs in multiple ways. For example, helper columns and their natural language descriptions illustrate how the system has implemented the solution, allowing users to verify that it matches their intent. Row summaries allow users to quickly identify rows that may need further inspection. Finally, expanding helper columns can enable users to make targeted repairs through \textit{programming-by-example} techniques such as FlashFill~\cite{DBLP:conf/popl/Gulwani11}.


\section{Principles of UX design for co-audit}\label{sec:principles}

Here is a list of design principles distilled from discussions at our workshop and afterward.
\begin{enumerate}
\item\label{princ:dont-rely-on-the-llm}
Don’t rely just on the LLM to co-audit itself.

It's an anti-pattern to trust the language model to audit itself.
For example, ChatGPT created a legal motion for a lawyer, but with made-up cases, rulings and quotes.
In an attempt at co-audit, the lawyer asked the program to verify that the cases were real.
ChatGPT responded incorrectly that the ``cases I provided are real and can be found in reputable legal databases''.
The lawyer submitted the brief to court and consequently was fined by the judge.\webcite{https://www.nytimes.com/2023/05/27/nyregion/avianca-airline-lawsuit-chatgpt.html}{Here’s What Happens When Your Lawyer Uses ChatGPT}{27 May 2023}
The example illustrates that co-audit may fail: an attempt at co-audit may or may not succeed in detecting an error.

\item\label{princ:ground-outputs-in-reliable-sources}
Ground outputs by citing reliable sources.

Citing sources is becoming a common pattern for AI-based chat systems\webcite{https://blogs.microsoft.com/blog/2023/02/07/reinventing-search-with-a-new-ai-powered-microsoft-bing-and-edge-your-copilot-for-the-web/}{Reinventing search with a new AI-powered Microsoft Bing and Edge}{7 February 2023}\:\webcite{https://blog.google/products/bard/google-bard-new-features-update-sept-2023/}{Bard can now connect to your Google apps and services}{19 September 2023}
and their plug-ins, such as the Wikipedia plug-in for ChatGPT.\webcite{https://diff.wikimedia.org/2023/07/13/exploring-paths-for-the-future-of-free-knowledge-new-wikipedia-chatgpt-plugin-leveraging-rich-media-social-apps-and-other-experiments/}{New Wikipedia ChatGPT plugin}{13 July 2023}


\item 
Teach the user how to construct effective prompts.

For example, apply grounded abstraction matching: after translating the user query into a system action, help the user understand how to consistently invoke the same system action through an editable example, e.g., a ``grounded utterance''
\cite{DBLP:conf/chi/LiuSNZ0T023}

\item\label{princ:visuals}
Inform the user with visuals as well as text.

For example, ConceptEVA provides co-audit of a document summary using techniques from exploratory visual analysis to show the concepts of interest underpinning the summary \cite{Zhang_2023}.

\item\label{princ:multiple-options}
Get the user to choose between multiple options.

Consider, for example, the LEAP environment for Python programming \cite{ferdowsi2023live}.
LEAP uses comments or code context to suggest multiple AI-generated suggestions. Its co-audit features include preview of each suggestion, and live editing of the suggestion including after its insertion into the program. 

\item\label{princ:most-likely-error}
Prioritize the user’s attention to most likely error.

For example, consider an experimental LLM-powered search experience \cite{spatharioti2023comparing} that uses colors to highlight high or low confidence measurements (numbers with units) as a co-audit feature for search results.
The highlighting relies on token probabilities from the LLM. The authors find overall that highlighting reduces the time taken for users to spot incorrect information.

\item\label{princ:decomposition}
Guide the user to attend to links between parts of
the output and the prompt or input document.

Take, for example, the nl2spec system \cite{cosler2023nl2spec}.
Given a natural description of a logical property, the system uses an LLM to synthesise a formula for the property in linear temporal logic. To aid co-audit, the system additionally uses the LLM to map subformulas of the output formula back to corresponding natural language fragments of the input.  Users can edit the subtranslations to correct errors.


\item
Involve two or more humans in a collaborative experience.

For example, one user may instigate the AI-generation of an output, while a different user may help to co-audit the output.
Here, we go beyond the prompt-response-audit cycle in Figure~\ref{fig:cycle}, which revolves around a single user, to be a collaboration amongst two or more humans, assisted by co-audit tools.


\item
Exploit tools designed for checking human-generated content.

In Section~\ref{sec:spreadsheets} we described a couple of research prototypes designed to co-audit AI-generated spreadsheet computations. Still, if the computation is represented as a formula, any existing debugger for a spreadsheet formula, such as FxD~\cite{drosos2023fxd}, can serve as a co-audit tool.

\item
Aim for positive “weekly cost-benefit” ratio for
user time invested.

This principle concerns measuring adoption of co-audit technology and using the measurement as a guide to ensure that co-audit is worth the time invested by users.
The principle is copied from a proposal for successful adoption of formal methods technology \citep{reid2020making}.
\end{enumerate}








\section{Potential pitfalls for co-audit: some research challenges}\label{sec:pitfalls}

In this section we consider potential risks arising from co-audit tools that can lead to attacks, system misuse (under-reliance or over-reliance) and even long-term societal harms. The content is derived from an internal workshop and represents the views of the participants. This section is not meant to provide an exhaustive overview, but we believe that starting to address the outlined issues is important for the co-audit experience. These risks are well known in the literature for related domains. We want to encourage the community to help us validate our hypothesis that they emerge for co-audit mechanisms as well, assess whether our proposed mitigations are valid and identify further gaps in our thinking.


\begin{description}
    \item[Security Attacks:] Security attacks are a common worry for AI-generated content and the Responsible AI practices should apply here, for instance to prevent excessive agency for co-audit tools. A recent report from OWASP\webcite{https://owasp.org/www-project-top-10-for-large-language-model-applications/}{OWASP Top 10 for Large Language Model Applications}{26 August 2023} details potential risks and mitigations that we should also consider. The use of co-audit may help mitigate potential security attacks (and other responsible AI concerns, such as privacy and fairness issues) by facilitating human oversight of the generated content for possible harms.  At the same time, the co-audit experience itself can be another possible point of attack for adversaries.  For example, attackers understanding the abilities of co-audit can formulate their attacks specifically to target co-audit weaknesses.  Also, the co-audit tool itself, if compromised, can be a rich target for exploitation as it has direct access to user information, the AI-generated content, etc. Further research is needed to identify attacks specific to co-audit and viable mitigations. 
    \item[System under-reliance or over-reliance:] We believe that co-auditing tools can be susceptible to both under-reliance and over-reliance.  While co-audit gives the user a greater understanding of the AI-generated content, it does not guarantee that the result is correct.   Likewise, if the cognitive load of co-auditing is too high, users could under-rely on the co-auditing system (for example, by having too many tasks to verify) or if they lack trust in the system. To minimise the cognitive load, the system should be non-intrusively integrated in the user's flow\adg{P2 keeping private for now "could include triage capabilities that draw attention to the most important issues"} and ask for very specific input from the human to make corrections. To improve trust, the system could show the work or rationale at various granularities including the impact of fixes (if we make a change at a certain point in the flow, be explicit about how the remainder of the flow gets impacted) and be explicit about the coverage of types of errors it can handle. Users could also over-rely on the co-auditing system if the system exhibits anthropomorphic behaviour (for example, by generating content claiming personified attributes) or if the remit and level of accuracy of the system are not clearly understood. Using linguistic cues to show uncertainty and articulating the specific assumptions made during the audit process could help the user understand the system's limitations. 
    \item[Societal harm:] There are significant long-term implications as humans increasingly leverage AI in their activities.  Co-audit, which provides users with greater confidence in their AI interactions, may help accelerate this trend.  How human/AI collaboration will evolve and what implications this evolution has for society are largely unknown. This uncertainty requires careful consideration of potential risks and unintended consequences that more powerful AI and better co-audit tools may enable.  For example, we could end up reducing creativity or diversity of thought by incentivising humans to produce outputs that co-audit systems can verify. There may also be
    an impact on workers
    (e.g., by turning every professional into their lesser-paid gig-based equivalent as users of co-audit tools),
    resulting in deskilling of workers, reduction of their economic value, and loss of jobs.
    In this scenario, there is a potential harm in the transfer of responsibility of review to the LLM with co-auditor as the LLM is given more and more responsibility as its capabilities increase.  There is also the risk of 
    a vicious cycle of skill-transfer
    from the human to the LLM as the LLM directly learns from successful human interactions and requires less and less guidance from human input over time.
    Moreover, the availability of AI and co-audit may result in humans losing their ability to effectively reason about knowledge sources and make judgements about goodness and correctness.  Use of AI and co-audit may lead to confusion  around who is responsible for an error that leads to harm.
    For co-auditing tools we believe that we need to have in place opt-in/opt-out functionality and sufficient customization such that the user can control the level of correctness, verbosity, and modality.
\end{description}

\section{Related work}\label{sec:related}

\subsection{Glassman's Conversational Model}
\citet{glassman2023designing} considers a conversational framework for human-AI communication, and links these to concepts from psychology. She identifies a series of points in the conversation including:
(a) the human forms an intent,
(b) expresses it as an utterance (akin to what we call a prompt);
(c) the AI executes inference,
(d) responds to the human;
(e) the human recognizes whether the AI's interpretation matches their intent,
(f) updates their mental model of the AI's capabilities,
(g) updates their mental model of when it can exercise those capabilities,
(h) confirms or refines their intent,
(i) revises the expression of their intent, and repeats from (b).

We can map our prompt-response-audit cycle in Section~\ref{sec:cycle} to her framework.
The cycle assumes a given human intent, which maps to Glassman's point (a).
The prompt step of the cycle maps to point (b).
The response step of the cycle maps to points (c) and (d).
The audit step of the cycle maps to points (e)-(i).
Of the possible decisions during audit, (A) and (B) map to (e), while (C) maps to (h), and the actions following the decisions map to (h) and (i).

So, co-audit amounts to any tool-supported human experience during audit, points (e)-(i). We have not emphasised that the human may update their mental model of the AI's capabilities, points (f) and (g), but that is worth considering during the design of co-audit tools.

\subsection{Some Examples and Non-Examples}\adg{maybe move earlier}

We use some examples to illustrate the concept of co-audit.

CheckList \cite{ribeiro-etal-2020-beyond} is a methodology with a companion interactive tool for behavioral testing of NLP models. CheckList is not intended as an end-user experience, but instead is aimed for use by providers of NLP systems for testing. Still, the user experience of CheckList appears to be a co-audit experience, where the tester double-checks the response from the model with tool support.

For a non-example of co-audit, consider CheckMate \cite{collins2023evaluating}, an interactive system to evaluate how effectively different language models can assist a human user in proving mathematical theorems. CheckMate offers a chat interface to different language models. It offers some simple features to make mathematical notation easier to read, such as rendering \LaTeX{} output. These cannot be described as enabling a systematic review of the output, so would not be considered co-audit. Still, we expect to see co-audit features in future copilots for mathematics.

\subsection{Related but distinct concepts}

Co-audit is distinct from \emph{algorithmic audit}, defined by \citet{DBLP:conf/fat/RajiSWMGHSTB20} as a systematic process managed by an organization to check that the creation and deployment of an AI system complies with its ethical expectations and standards.
\adg{P2 have earlier citation for algorithmic audit}
\citet{HAI_LLM} argue for the importance of rigorous algorithmic audit of LLMs before deployment. They report recent success in algorithmic audit of LLMs by using LLMs themselves, and highlight the importance of sensemaking and human-AI communication to leverage complementary strengths of humans and generative models in collaborative auditing of LLM output.

Co-audit is also distinct from \emph{evaluation} of an NLP system's performance as a whole, for instance, on a suite of benchmarks.  See the classic review by \citet{Jones93}, for example.  
The distinction is that co-audit is for a specific instance of a user interaction, while algorithmic audit or NLP evaluation consider general properties of the whole system or model, rather than individual actions.
Consider also mechanisms to classify between human- and AI-generated content \cite{dhaini2023detecting,kirchenbauer2023reliability}. These mechanisms are not co-audit either, because co-audit starts from a piece of AI-generated content.

\citet{co-piloted-auditing} examine \emph{co-piloted auditing}, where the user's intent is to use foundation models to perform a financial audit---a different usage of the word than our usage for a systematic review of an AI response, whether or not in the financial domain. They report numerous prompts to perform financial auditing.


An alternative definition of co-audit might be that co-audit is a human and AI auditing their work together. This metaphor presents human and AI as two entities who, while collaborating to produce an output, audit the output together. Indeed, such anthropomorphic metaphors are common, but the practice may be misleading \citep{SEF2020}.
Hence, we prefer to frame co-audit in terms of tool use by a human user, instead of a metaphorical collaboration \citep{collaborative23,abercrombie2023mirages}.


\subsection{Mistakes}
The basic idea that AI makes mistakes, and humans therefore need support to discover and deal with those mistakes, is not in itself new. 
%
Dealing with errors has been a key part of research on human-AI interaction; for instance, several of the guidelines for human-AI interaction \cite{DBLP:conf/chi/AmershiWVFNCSIB19} concern how to deal with AI errors. \citet{liao2023ai} identify research challenges raised by LLMs for AI transparency, citing research on explanations of individual model outputs 
\citep{DBLP:conf/kdd/Ribeiro0G16,DBLP:conf/icml/KohL17,DBLP:conf/nips/LundbergL17,DBLP:conf/fat/Russell19,DBLP:conf/fat/UstunSL19}.

The term ``explanation'' encompasses a wide variety of techniques to assist in the user experience of AI systems, and subsumes aspects of the experience that relate to the correctness of model outputs, such as evaluating model outputs and then correcting them. Research in explainable AI has thus long encompassed the specific subset of concerns we delineate here as the need for co-audit. For example, Lim and Dey's intelligibility types for context-aware applications \cite{lim2009assessing} include types of information that help users detect errors (e.g., why, how, why not, what if, certainty), as well as correct them (control). In Tintarev and Masthoff's framework of criteria for explanations \cite{tintarev2010designing}, co-audit is covered by a combination of transparency (explain how the system works) and scrutability (allow users to tell the system it is wrong).

\adg{The phrase "co-audit for copilots" could apply here}

While the concerns of co-audit are not new, the broad introduction of generative AI with its particular strengths and limitations is accompanied by a number of shifts in the nature of the user experience of intelligent systems. The shift in user experience has been described as the ``generative shift'' \cite{sarkar2023code}. In particular, users will apply AI more intensively to existing tasks, they will apply it more extensively to tasks which they previously did not apply AI to, and they will apply it more frequently than before. 

There are concomitant shifts in the user experience of \emph{error}. System errors will have the opportunity to become more consequential, span across a wider set of tasks, and occur more frequently than before. In earlier generations of interactive machine learning (IML) tools, the user was responsible for directly training the model to shape its behaviour and make it perform more optimally in the long run, e.g., the user of a photo editing tool training an image segmentation engine to remove the background from their images \cite{fails2003interactive}, a spreadsheet user building a model of their data to impute missing values \cite{gordon2015probabilistic,sarkar2015interactive}, or an email client user training a spam filter \cite{kulesza2015principles}. In contrast, the prevailing commercial idioms of user experience for contemporary generative AI tools are far more passive: users can control the behaviour of the model in individual instances through their prompts, but they cannot participate in its construction on the more fundamental level that was common in earlier IML tools, such as through the curation of the training set, provision of reinforcement learning feedback signals, or adjustment of training hyperparameters. Consequently, the interaction context inhabited by Lim and Dey's ``control'', and Tintarev and Masthoff's ``scrutability'', is increasingly divergent from the interaction context inhabited by generative AI tools; we thus offer co-audit as a practical reinterpretation of those concepts.\adg{P2 Maybe use the word "incommensurable" as used by Kuhn}


\section{Conclusion and call to action}\label{sec:conc}



Just as individuals learned how to best interact with search engines over a period of years, they will have to learn how best to interact with generative AI and copilots. Human skills in both prompt engineering and audit are needed. We have discussed a range of aspects of co-audit tools to help and teach people to audit AI outputs. We believe that some of these co-audit tools and skills should exist across copilot experiences while other skills will be domain-specific.  Our goal is to jump-start research into these co-audit experiences to maximize the benefits of using AI while minimizing risks. As a starting point, we have outlined principles of UX design for co-audit and provided examples to illustrate how these principles can be applied in practice.  But many questions and important research challenges remain.  Co-audit experiences will evolve as the underlying foundation models and copilot infrastructures will evolve.  But because co-audit directly connects with user, it is especially important that these experiences remain similar even as the underlying technology changes.  To ensure that we create enduring co-audit experiences, we challenge the research community to address the following questions.

\begin{itemize}
\item How to integrate a co-audit user experience into copilot chats?
\item How to integrate the co-audit experience across applications that relate to the same content?
\item How can co-audit help individuals create, maintain, and check increasing volumes of content that AI generation enables? 
\item What co-audit skills are transferable across different diverse vertical uses of AI?  
\item Will the principles of understanding the co-audit experience remain stable as AI evolves or will it require frequent redesign?
\end{itemize}

To end, recall that Hal Varian, chief economist at Google, famously said that dull old statistics would be the cool skill of the 2010s, the decade of big data \citep{sexiestjob2012}.
The 2020s are the decade of generative AI.
We say that dull old auditing will be the cool skill of the 2020s.
And we call for research on co-audit experiences to help more people thrive this decade and beyond.



\subsection*{Acknowledgements}
Thanks to Ashley Llorens, Brittney Muller, and all at Microsoft Research Outreach for supporting the workshop reported in this paper.
Gavin Abercrombie provided technical assistance.
Christian Canton offered useful advice and suggestions.
We are grateful to Peter Lee for his feedback and support, and to Kevin Scott for announcing our research on co-audit in the Verge.\webcite{https://www.theverge.com/23733388/microsoft-kevin-scott-open-ai-chat-gpt-bing-github-word-excel-outlook-copilots-sydney}{Microsoft CTO Kevin Scott thinks Sydney might make a comeback}{23 May 2023}
And finally, we acknowledge Bill Gates for his feedback and encouragement.

\bibliography{co-audit}
\end{document}